\documentclass[preprint,10pt]{aastex}

\newcounter{affil}

\newcommand{\bfR}{\ensuremath{\vec{R}}}
\newcommand{\bfr}{\ensuremath{\vec{r}}}
\newcommand{\nbar}{\ensuremath{\bar{n}}}
\newcommand{\Var}{{V}}
\newcommand{\Rmax}{{R_{\rm max}}}
\newcommand{\xiis}{{\xi_{\rm is}}}
\newcommand{\wis}{{w_{\rm is}}}

\newcommand{\beq}{\begin{equation}}
\newcommand{\eeq}{\end{equation}}
\newcommand{\beqa}{\begin{eqnarray}}
\newcommand{\eeqa}{\end{eqnarray}}

\begin{document}

\lefthead{DEPROJECTED DENSITIES}
\righthead{EISENSTEIN}

\title{Deprojecting Densities from Angular Cross-Correlations}
\author{Daniel J.\ Eisenstein}   
\affil{Steward Observatory, University of Arizona, 
		933 N. Cherry Ave., Tucson, AZ 85721 \\
		deisenstein@as.arizona.edu}

\begin{abstract}
I present a model-independent spherically symmetric density estimator
to be used in the cross-correlation of imaging catalogs with objects
of known redshift.  The estimator is a simple modification of the
usual projected density estimator, with weightings that produce a
spherical aperture rather than a cylindrical one.
\end{abstract}


\section{Introduction}

Measuring galaxy properties as a function of their local environment
is a central task of observational extragalactic astronomy.  
Doing so requires a measure of density.  With redshift surveys,
one can estimate densities by counting other spectroscopic galaxies in a 
redshift-space window around each primary object.  
However, imaging catalogs are generally much
deeper than spectroscopic catalogs, which suggests that cross-correlating
the imaging catalog with the spectroscopic catalog should yield 
useful densities even for the faintest spectroscopic objects.

The presence of at least one spectroscopic redshift in each pair is
of great utility because it means that one can immediately map angular
separations to physical distances and derive intrinsic properties
(i.e.~luminosities) for both the spectroscopic object and the correlated
objects from the imaging catalog.  This opportunity has been used
extensively in the early study of galaxy clustering 
\citep{Dav78,Yee87,Lil88,Sau92} 
and in the study of dwarf galaxies in groups 
\citep[e.g.,][]{Fer91} and in the field \citep{Phi87,Vad91,Lor94,Lov97}.  
Of course, the correlated objects are still seen in projection, with a mix
of spatial separations at each transverse separation.

Angular correlations can be inverted to spatial correlations,
and hence to a form of density,
by the assumption of isotropy, and many of the above studies 
did this
by assuming that the correlation functions are power laws in scale
\citep[e.g.,][]{Phi85}.
\citet{Sau92} and \citet{Lov97} go one step further by using the
assumed power-law as an optimal filter.  \citet{Fal77} suggest
that the general inversion can be done with a smoothing filter.
\citet{Bau93} use a regularized Lucy's iteration to do a similar 
inversion, while \citet{Dod00} and \citet{Eis01} use other 
smoothing priors.

In the study of galaxy environmental dependences on small scales,
it is useful to pursue a more model-independent measure of the density.
On scales of 1 Mpc, it is quite possible that the correlation 
functions will deviate from power laws (and certainly from a 
uniform power law) in ways that depend on luminosity, star-formation
rate, or other variables \citep[e.g.,][]{Scr02}.

Here I describe a method to recover 
density estimates in spherically symmetric 
real-space windows from the cross-correlation of imaging and
spectroscopic catalogs independently of the shape of the correlation
function.  The resulting formulae are a simple
alteration of the usual background subtraction methods.  The
method can be subdivided so as to yield noisy density estimates
for individual spectroscopic objects that can then be averaged
according to whatever subclasses one might desire.

\section{Angular Cross-correlations}

\subsection{Definitions}

We begin with two samples of objects, one with spectroscopic
redshifts and the other without.  We bin the spectroscopic sample
into thin redshift bins (indeed, we can consider each spectroscopic
object separately) and select a subsample of the imaging catalog
using the known redshift.  For example, one might select a sample
in a particular luminosity range (i.e. the objects would have the
desired luminosity were they at the spectroscopic redshift).
We adopt a flat-sky coordinate system in which the transverse directions
are measured as distance $\bfR$ at the given redshift and the linear 
direction is measured as distance $Z$ where $Z=0$ is at the redshift.
We write the three-dimensional position $(\bfR,Z)$ as $\bfr$.

Of course, the imaging subsample samples a range of $Z$, not just
$Z=0$.  We describe the homogeneous density of the sample as 
$\phi(Z)$, where this is the number per unit $Z$ and per unit 
transverse area.  The areal number density is 
\beq
\nbar=\int_{-\infty}^\infty dZ\;\phi(Z).
\eeq

If the true spatial cross-correlation between the spectroscopic
object and the imaging subsample is $\xiis(r)$, then the 
angular correlation is
\beq
\wis(R) = {1\over\nbar}\int_{-\infty}^{\infty}dZ\;\phi(Z)\xiis(\sqrt{Z^2+R^2}) 
\approx {2\phi_0\over \nbar}\int_0^\infty dZ\;\xiis(\sqrt{Z^2+R^2})
\label{eq:wdef}
\eeq
where the last equality assumes that the selection function $\phi$
is essentially constant over the scale on which $\xiis$ is not negligible.
We also assume that the angular diameter distance is constant in this region.
We denote $\phi(0)$ as $\phi_0$.
Equation (\ref{eq:wdef}) is a simple form of Limber's equation
\citep{Lim53,Gro77}.

\subsection{Deprojection}

It is well-known that equation (\ref{eq:wdef}) can be inverted
as an Abel integral \citep{von08}, so that
\beq\label{eq:abel}
\xiis(r) = -{\nbar\over \phi_0\pi}\int_r^\infty {dR\over\sqrt{R^2-r^2}}{d\wis(R)\over dR}.
\eeq
However, because this involves a derivative of the measured $\wis(R)$,
it is noisy.  

We address this by measuring only an integral of $\xiis$
\beq\label{eq:Deltadef}
\Delta = {1\over V}\int_0^\infty 4\pi r^2dr\;\xiis(r) W(r)
\eeq
where $V = \int_0^\infty 4\pi r^2dr\;W(r)$.  $W(r)$ is our smoothing
window.  $\Delta$ has a very useful physical meaning: it is the
average overdensity of objects from the imaging catalog in the 
neighborhood (as defined by $W$) of a spectroscopic object.
Note that $\Delta$ is defined in 3-space with a spherically
symmetric window; it is {\it not} a projected quantity.

Inserting (\ref{eq:abel}) into (\ref{eq:Deltadef}) and switching
the limits of integration yields
\beq
\Delta = -{\nbar\over \phi_0V\pi}\int_0^\infty dR\;{d\wis\over dR}
\int_0^R dr\;{4\pi r^2 W(r)\over\sqrt{R^2-r^2}} 
= -{2\pi \nbar\over\phi_0V}\int_0^\infty dR\;{d\wis\over dR} F(R)
\eeq
where 
\beq\label{eq:Fdef}
F(R) = {2\over \pi}\int_0^R dr\;{r^2 W(r)\over \sqrt{R^2-r^2}}.
\eeq
For bounded $W(r)$, $F(0)=0$ and $F(R)\rightarrow V/2\pi^2R$ for
large $R$.
Integrating by parts yields
\beqa\label{eq:Delta}
\Delta &=& {\nbar\over \phi_0V} \int_0^\infty 2\pi R\,dR\;\wis(R) G(R) \\
G(R) &\equiv& {1\over R}{dF\over dR}.
\eeqa
A constant $\wis$ integrates to $\Delta=0$, so adding a constant
to $\wis$ doesn't change $\Delta$.  $G(R)\rightarrow -V/2\pi^2R^3$ for
large $R$.

Now let us consider our measurement of $\wis(R)$.  In a small radial
bin from $R$ to $R+dR$, we would estimate $1+\wis(R)$ as the ratio of
the observed counts of pairs in that radial range to the expected
number.  The expected number is $2\pi \nbar R\,dR$.  Treating the
integral as a Riemann sum in which the bins are so small as to 
contain 0 or 1 observed pair leads to the conclusion that
\beq\label{eq:Deltasum}
\Delta = {1\over N_{sp}}\sum_{j\in\{sp\}} {1\over \phi_0V}\sum_{k\in\{im\}} 
G(R_{jk})
\eeq
where the sums are over the objects in spectroscopic subsample 
and the imaging subsample, respectively, 
$N_{sp}$ is the number of spectroscopic objects,
and $R_{jk}$ is the transverse separation of the $j^{th}$ spectroscopic object
to the $k^{th}$ imaging object.

An important point is that we can now treat each spectroscopic object
separately, yielding the following noisy measure of the overdensity around
object $j$:
\beq\label{eq:Deltaj}
\Delta_j = {1\over \phi_0V}\sum_{k\in\{im\}} G(R_{jk}).
\eeq
Note how simple this formula is: one 
counts the imaging objects, weighting by $G(R)$, and divides
by the expected number of objects in the real-space window ($V\phi_0$).  
We can recover the average
density around any subset of the spectroscopic sample simply
by averaging the selected $\Delta_j$.

It is interesting to compare equation (\ref{eq:Deltaj}) 
to a more conventional background-subtraction
method in which one sums all of the objects in an angular aperture and
subtracts an appropriately scaled value of the areal density averaged
over the entire survey.  This would correspond to a $G$ function that
was constant and positive for $R$ less than the aperture and then
constant and negative for all greater $R$.  The difference is that
this background subtraction would give an estimate for the density
in a cylindrical region, in which the axis of the cylinder lies along
the line of sight and is much longer than the radius of the cylinder.
The resulting density would sample the correlation function $\xiis$ at
a wide range of radii.  The formula given here creates a compact region
in all three dimensions.  Some workers \citep[e.g.,][]{Gai97,Val97} have used annular
regions for the determination of the background.  This truncates the
cylindrical region in some fashion, but the detailed effects were not assessed.

\subsection{Gaussian Windows}

For a useful and illustrative example, we will treat the case in which 
the window is assumed to be a Gaussian
$W(r) = \exp(-r^2/2a^2)$.  The volume of the window is 
$V=(2\pi)^{3/2}a^3$.  Then
\beq
F(R) = {2\over\pi}\int_0^R dr\;{r^2 e^{-r^2/2a^2}\over \sqrt{R^2-r^2}} 
= {R^2\over 2} e^{-s}\left[I_0(s)-I_1(s)\right]
\eeq
where $s=R^2/4a^2$ and $I_n$ are modified Bessel functions of the 
first kind \citep[3.364.1]{GR}.  Then
\beq\label{eq:Ggauss}
G(R) = e^{-s}\left[I_0(s)-2sI_0(s)+2sI_1(s)\right].
\eeq

The weighting function $G$ is shown in Figure 1.  The function is
smooth, with a positive peak at $R\approx a$ and a broader negative peak
at $R\approx 3a$.  As expected, to estimate the density averaged over
the window $W$, one is counting the nearby objects and subtracting
a count of objects slightly further away.

\subsection{Boundaries and Masks}

Equation (\ref{eq:Delta}) requires an integration over all radii and
since $G\propto R^{-3}$ at large radius, this integration converges
as $\sim 1/R$.  
If one splits the integral in equation (\ref{eq:Delta}) at some 
radius $\Rmax$, inside of which one will do the counting in equation
(\ref{eq:Deltasum}), then the residual error from larger radii is
\beq\label{eq:DeltaErr}
\Delta_{res} = {\nbar\over \phi_0V} \int_\Rmax^\infty 2\pi R\,dR\;[1+\wis(R)] G(R) 
\eeq
Here, we have to include the constant background term because it
will no longer integrate to zero.
We can estimate the correlated term by treating $\wis$ as a power-law 
$\wis(R) = \wis(\Rmax) (\Rmax/R)^\alpha$, 
where $\wis(\Rmax)$ is the value at the truncation radius and 
$\alpha$ is the slope of the power-law, typically 0.7--0.8 in past
measurements.  Then, using $G(R)\approx -V/2\pi^2R^3$ in the second term, 
we find
\beq
\Delta_{res} = -{\nbar\over \phi_0}\left[
{2\pi F(\Rmax)\over V} + {\wis(\Rmax)\over \pi\Rmax}{1\over 1+\alpha}
\right].
\eeq
Taking $\xiis\propto r^{-1-\alpha}$, we have
\beq
{\nbar\over\phi_0}\wis(R) = R\xiis(R){\Gamma({\alpha\over2})\sqrt\pi\over \Gamma({\alpha+1\over2})}
\eeq
One finds that 
$\Delta_{res} = -2\pi\nbar F(\Rmax)/\phi_0V-0.70\xiis(\Rmax)$ 
for $\alpha=0.75$.  
Applying this as a correction does introduce some model dependence
on the form of $\wis$, but one can pick $\Rmax$ so that the correction
is rather small.  In this way, one can avoid summing over all pairs
of spectroscopic and imaging objects.

The circular region $R<\Rmax$ may still include regions that are
outside the survey or its mask.  If one writes $\Phi(R)$ as the
fraction of the annulus of radius $R$ that is within the survey, then one can 
weight the counts in equation (\ref{eq:Deltasum}) by $1/\Phi(R)$.
However, $\Phi(R)$ may be expensive to compute for each spectroscopic 
galaxy.  An alternative
method is to generate a catalog of random points that are uniformly 
distributed {\it outside} of the survey region and then add to
equation (\ref{eq:Deltaj}) the sum over
those points closer than $\Rmax$, weighting by $G(R)[1+\wis(R)]$ and
the ratio of $\nbar$ to the random catalog surface density.
This effectively interpolates over the masked regions.  If one 
treats $\wis$ as a function of fixed shape but unknown amplitude, then one can do the sum of the clustered
term separately and renormalize after one has determined
the amplitude from the co-added $\Delta_j$.  Of course, since one
is in some fashion assuming the answer, one should exclude
spectroscopic galaxies for which the masked contribution is
large.

\subsection{Errors and Optimal Weighting}

When averaging the individual estimates $\Delta_j$ from a sample
of spectroscopic objects, one can achieve more precision by using 
non-uniform weighting in the mean.
The optimal weighting is inverse variance; however, one should
remember that approximations to inverse variance may be much simpler
to compute and yet still close to optimal.  One should never weight
by variances that are derived from single data points, e.g.\ the 
square root of the observed number of companions to a given galaxy,
because this will bias the resulting mean.

The statistical variance of the $\Delta_j$ estimator comes from two sources,
the clustering of the galaxies and shot noise.
The clustering term involves the three-point correlation function,
with one galaxy from the spectroscopic sample and two from the 
imaging sample.  If the first of these is at the origin and the
latter two are at $\bfr_1$ and $\bfr_2$, then we denote the three-point
correlation function as $\zeta(\bfr_1,\bfr_2)$.  The expected densities
at two points given a spectroscopic galaxy at the origin is \citep{Pee80}
\beq\label{eq:rhoexp}
\left<\rho(\bfr_1)\rho(\bfr_2)\right> = \phi(Z_1)\phi(Z_2) 
\left[1+\xiis(r_1)+\xiis(r_2)+
\xi_{ii}(|\bfr_1-\bfr_2|)+\zeta(\bfr_1,\bfr_2)\right] 
\eeq
where $\xiis$ is the two-point correlation between imaging and 
spectroscopic galaxies and 
$\xi_{ii}$ is the correlation of two imaging galaxies.

The clustering contribution to the variance of $\Delta_j$ is then
\beqa\label{eq:VarCl}
\Var_{\rm cl}(\Delta_j) &=& 
\int d^2R_1 \int d^2R_2 G(R_1) G(R_2) \int dZ_1 dZ_2
\times \nonumber \\
&& \quad 
\left[\xi_{ii}(|\bfr_1-\bfr_2|)+\zeta(\bfr_1,\bfr_2) - \xiis(r_1)\xiis(r_2)\right]
{\phi(Z_1)\phi(Z_2)\over \phi_0^2V^2}.
\eeqa
The last term is simply $\left<\Delta_j\right>^2$.
The terms in $\xiis$ in equation (\ref{eq:rhoexp}) integrate to zero.
The $\zeta$ term involves all three objects, so at the level of approximation
in equation (\ref{eq:wdef}), we can assume $\phi(Z_1)=\phi(Z_2)=\phi_0$.
However, the $\xi_{ii}$ term represents correlated imaging galaxies that are 
uncorrelated with the spectroscopic object, so they may have $Z$
far from 0 and $\phi(Z)\ne\phi_0$.  
Integrating over $Z$ will yield the angular correlation of
the imaging catalog.  Indeed, for $Z\ne0$, the angular separations implicit
in our definition of the transverse coordinate $\bfR$ correspond to different
physical scales.  Doing this correctly again yields the angular
correlation function.
We thus simplify equation (\ref{eq:VarCl}) to 
\beqa
\Var_{\rm cl}(\Delta_j) &=& \Var_{\rm 2pt} + \Var_{\rm 3pt}\\
\Var_{\rm 2pt} &\equiv& \left(\nbar\over\phi_0V\right)^2
\int d^2R_1\int d^2R_2 G(R_1) G(R_2) w_{ii}(|\bfR_1-\bfR_2|)
\nonumber\\
\Var_{\rm 3pt} &\equiv&
\int {d^3r_1\over V} \int {d^3r_2\over V} G(R_1)G(R_2)\left[\zeta(\bfr_1,\bfr_2)-\xiis(r_1)\xiis(r_2)\right] \nonumber
\eeqa
The first term can be done quickly by Fourier methods, as can
the second if one adopts the hierarchical ansatz \citep{Gro77}
to write 
$\zeta(\bfr_1,\bfr_2) = Q\{\xiis(r_1)\xiis(r_2)+[\xiis(r_1)+\xiis(r_2)]\xi_{ii}(|\bfr_1-\bfr_2)|)\}$.
The two-dimensional Fourier transform of $G(R)$ is 
\beq
\int d^2R\;e^{i\vec{k}\cdot\bfR} G(R)
= 4\int_0^\infty dr\;r \sin(kr)W(r)
\eeq
Note that while $\Var_{\rm 3pt}$ is a contribution to the variance
about the mean $\Delta_j$, it is not necessarily noise!  Much of it {\it is}
the density around the particular object, which of course has scatter
from the mean.

The shot noise or Poisson contribution to the variance is based on
the expected counts, including clustering, which are $\nbar[1+w_{is}(R)]$.
The variance is
\beq
\Var_{\rm sn}(\Delta_j) = \int d^2R\;\left[G(R)\over \phi_0V\right]^2 \nbar[1+\wis(R)].
\eeq
We refer to these two terms as the homogeneous and clustered shot noise,
respectively.

\begin{figure}[tb]
\plotone{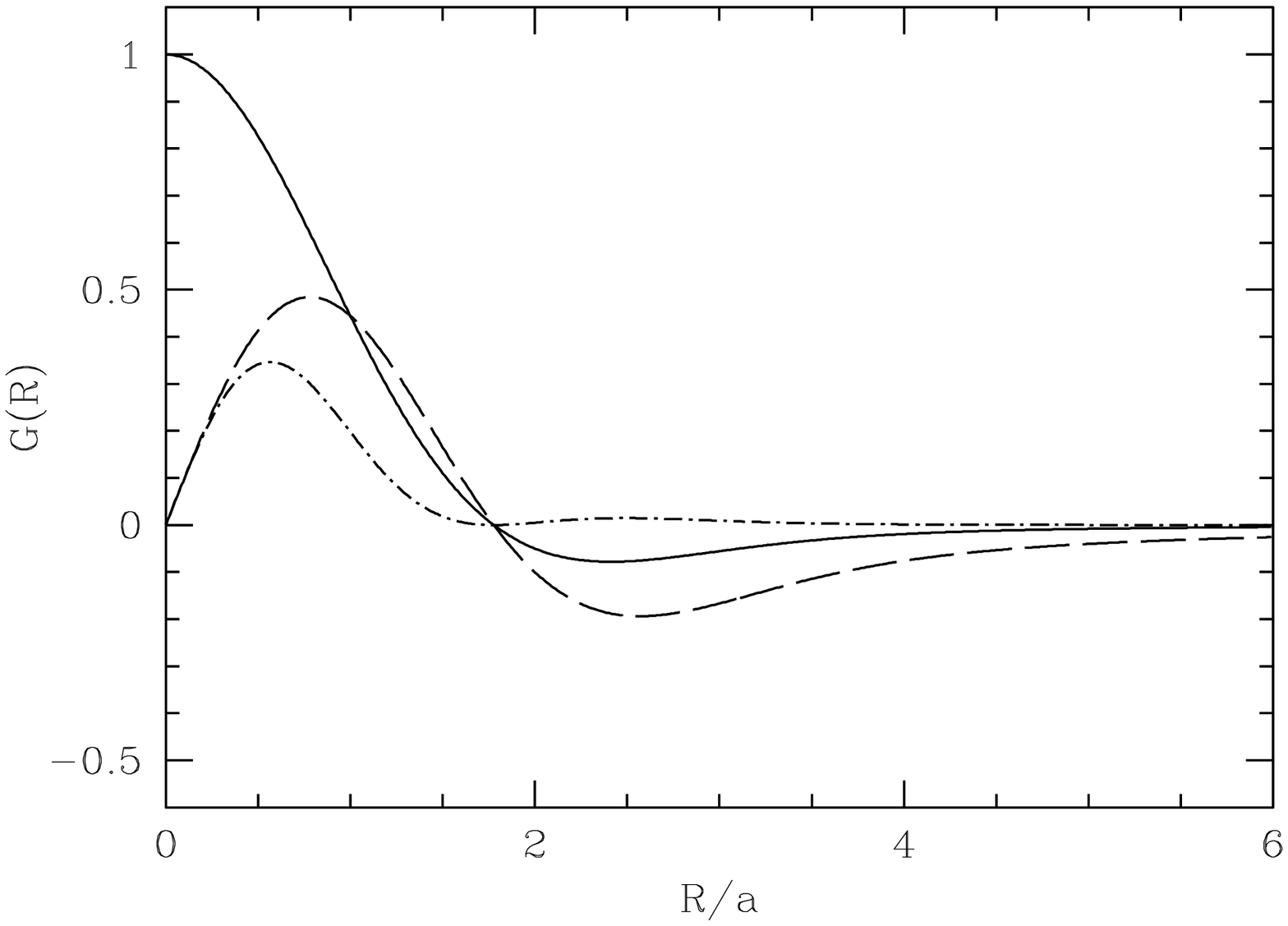}
\caption{\label{fig:g}
({\it solid line}) The integration kernel $G(R)$ for 
the Gaussian window.
({\it dashed line}) $R G(R)$, which is the weight per radial bin
for a uniform background.
({\it dot-dashed line}) $R G^2(R)$, which is the shot noise variance 
per radial bin for a uniform distribution.
}
\end{figure}

For the Gaussian window (eq.~[\ref{eq:Ggauss}]), the homogenous shot noise
becomes $2a^2\nbar/(V\phi_0)^2$.  
Figure \ref{fig:g} shows the contribution per
radial bin to the variance in the shot noise.  Essentially all of the
shot noise arises at $R<1.5a$; in other words, the fact that one 
is subtracting the background with a region at moderate radius rather
than the entire sample adds little extra noise to the density estimator.

The four above contributions---$\Var_{\rm 2pt}$, $\Var_{\rm 3pt}$,
homogeneous shot noise, and clustered shot noise---all have different
scalings with the depth of the survey, the size of the window, and
the clustering strength.  For a typical survey thickness $L\equiv \nbar/\phi_0$
and a typical window radius $a$,
the contributions to the variance are roughly $(L/20a)\Delta$,
$\Delta^2$, $(L/125a)(1/a^3\phi_0)$, and $(\Delta/32)(1/a^3\phi_0)$,
respectively.  The numerical coefficients are for illustration 
only\footnote{The numerical factors are computed for
the Gaussian window with the assumptions of a power-law correlation function
scaling as $r^{-1.8}$ with uniform bias for all galaxies,
an $\alpha=-0.9$ Schechter luminosity function
with imaging catalog luminosity cuts between $0.4L^*$ and $2.5L^*$,
the hierarchical form of $\zeta$ with $Q=1.3$,
and a non-expanding cosmology (Euclidean metric, no K-corrections).}.
The 2-point clustering term dominates on large scales; the clustered
shot-noise on small scales.

The process of averaging many $\Delta_j$ into a mean $\Delta$
introduces additional error terms from the correlations of the
spectroscopic galaxy positions, including contributions from
the four-point correlation function.  This is not surprising 
because $\Delta$ is an integral of $\wis(R)$, whose covariance normally 
involves the four-point function.  Neglecting these new terms in favor of the 
equations above corresponds to the assumption that the shot noise of
the spectroscopic sample exceeds its clustering \citep[see the 
expansions of][]{Ber94,Ham97}.  This is often a good assumption on small
scales, particularly if one is considering only a small subset of 
the spectroscopic sample.

A crucial assumption of the analysis is that the objects that are
uncorrelated but in close projection with the spectroscopic object
are statistically identical to those in other parts of the sky.
Magnification from weak lensing can violate this assumption in principle
\citep[e.g.,][]{Val97}.  
Moreover, selection biases in cluster catalogs owing to superposition 
of unrelated structures \citep[e.g.,][]{Val01} are not reduced by this method.

\subsection{From Kernels to Windows}

If we are given $G(R)$, we can use the Abel integral to find the
corresponding $W(r)$:
\beq\label{eq:WfromG}
W(r) = {1\over r}\int_0^r dR{R G(R)\over \sqrt{r^2-R^2}}.
\eeq
If $W(r)$ is to be bounded and $V\ne0$, $G(R)$ must have an asymptotic 
form of $-V/2\pi^2R^3$.  In particular, this means that one cannot
have $G=0$ for all large radii, thereby avoiding summations over large
pair separations.

\subsection{Other Windows}

As one might expect, the spherical tophat is easily calculated 
but poorly behaved.  Taking $W(r)=1$ for $r<b$, we find 
\beq
G(R) = \left\{\begin{array}{lc}\displaystyle
1 & R<b \\
\displaystyle
{2\over\pi}\left[\sin^{-1}{b\over R} - {b\over \sqrt{R^2-b^2}}\right]
& R>b
\end{array}\right.
\eeq
This diverges as $R\rightarrow b$ from above.  The singularity
is integrable in $\Delta$ but divergent in the shot noise, making
the tophat a poor choice.

Another simple choice that retains the advantage that $W=0$ for $r>b$
is $W(r)=1-r^2/b^2$ for $r<b$.
We find 
\beq
F(R) = \left\{\begin{array}{lc}\displaystyle
{R^2\over 2} - {3R^4\over 8b^2} & R<b \\
\displaystyle
{2\over\pi}\left[\left(-{b\over 4}+{3R^2\over 8b}\right)\sqrt{R^2-b^2}
+ \left({R^2\over 2} - {3R^4\over 8b^2}\right)\sin^{-1}{b\over R}\right]
& R>b
\end{array}\right.
\eeq
and
\beq\label{eq:gpoly}
G(R) = \left\{\begin{array}{lc}\displaystyle
1-{3R^2\over 2b^2} & R<b \\
\displaystyle
{2\over \pi}\left[ {3\sqrt{R^2-b^2}\over 2b}+
\left(1-{3R^2\over 2b^2}\right)
\sin^{-1}{b\over R}\right]
& R>b
\end{array}\right.
\eeq
This is continuous at $R=b$ but has somewhat worse shot noise
properties than the Gaussian window.

\section{Conclusions}

I have presented a model-independent estimator for the spherically
averaged overdensity of imaging catalog objects around spectroscopic 
objects.  The method is simple to apply; one can view it as an 
alteration of standard background subtraction methods to yield 
spherical apertures rather than cylindrical ones.  The method
does not require an explicit inversion of the angular cross-correlation 
function to a spatial correlation function, although clearly if one finds 
oneself measuring the densities in multiple apertures of different sizes, 
one is effectively reverting back to an inversion method.  
The primary advantages of the new method compared to integrating
the output of an inversion (i.e., computing eq.~[\ref{eq:abel}] and
then eq.~[\ref{eq:Deltadef}]) are that one does not need a smoothing
prior and that the statistic can be applied to each spectroscopic
galaxy independently (eq.~[\ref{eq:Deltaj}]),
leaving one free to sum the results {\it post facto} across as many 
spectroscopic subsamples as one desires.
Like other angular methods, 
the new density estimator is unaffected by redshift distortions, which gives it
an advantage on small scales over density estimation from spectroscopic 
catalogs.

Past work has assumed power-law correlation functions to recover
spatial densities.  This allows one to achieve higher signal-to-noise
ratio and hence is a better choice for some applications.  However,
when probing the dependence of galaxy properties on small-scale
environment, the model independence of the density estimator presented
here is a valuable advantage.  With today's large surveys
\citep[e.g.,][]{Yor00,2dF}, statistical precision is sometimes
less precious than systematic control.

While it is clear that one can profitably consider the dependence
of average density on the properties of the spectroscopic objects
\citep{Hog02}, it is worth pointing out that one can also derive
densities for subdivisions of the imaging catalog.  For example,
one can probe the color and/or luminosity distribution of objects
within a spherical aperture of a particular set of spectroscopic
objects (e.g., galaxies or clusters).
A speculative application would be to couple this approach to 
galaxy-galaxy weak lensing mass estimates.  If one had a mass
estimate for each imaging object, assuming the spectroscopic
redshift, then one could find the masses of objects that are
correlated with the spectroscopic tracer.

\acknowledgements 
I thank David Hogg, Jon Loveday, Tim McKay, Ann Zabludoff, and Dennis Zaritsky 
for useful discussions.
D.J.E. was supported
by National Science Foundation (NSF) grant AST-0098577
and by a Alfred P. Sloan Research Fellowship.

\clearpage

\end{document}